\documentstyle[12pt]{article}
\newcommand{\beql}{\begin{eqnarray}}
\newcommand{\eeql}{\end{eqnarray}}
\newcommand{\beq}{\begin{equation}}
\newcommand{\eeq}{\end{equation}}
\newcommand{\bc}{\begin{center}}
\newcommand{\ec}{\end{center}}
\newcommand{\bfr}{\begin{flushright}}
\newcommand{\efr}{\end{flushright}}
\newcommand{\bfl}{\begin{flushleft}}
\newcommand{\efl}{\end{flushleft}}
\newcommand{\bl}{\begin{large}}
\newcommand{\el}{\end{large}}
\newcommand{\bll}{\begin{Large}}
\renewcommand{\ell}{\end{Large}}
\newcommand{\blll}{\begin{LARGE}}
\newcommand{\elll}{\end{LARGE}}
\newcommand{\bdes}{\begin{description}}
\newcommand{\edes}{\end{description}}
\newcommand{\bitem}{\begin{itemize}}
\newcommand{\eitem}{\end{itemize}}

\newcommand{\benum}{\begin{enumerate}}
\newcommand{\eenum}{\end{enumerate}}

\newcommand{\vsl}{\vspace{1cm}}

\newcommand{\blarge}{\begin{Large}}
\newcommand{\elarge}{\end{Large}}
\def\lromn#1{\uppercase\expandafter{\romannumeral#1}}

\def\blist{\begin{list}{\setlength{\rightmargin}{\leftmargin}}}
\def\elist{\end{list}}
\begin{document}

\input epsf

\begin{titlepage}

\begin{flushright}
\begin{large}
TU/93/445\\
September 1993
\end{large}
\end{flushright}

\vspace{12pt}

\begin{center}
\begin{Large}

\renewcommand{\thefootnote}{\fnsymbol{footnote}}
\bf{End Point of Hawking Evaporation \\
-- Case of Integrable Model --}\footnote[1]
{Work supported in part by the Grant-in-Aid for Science Research from
the Ministry of \\ \hspace*{0.6cm} Education, Science and Culture of
Japan No. 00040633 and No. 03640257}

\end{Large}

\vspace{36pt}

\begin{large}
\renewcommand{\thefootnote}{\fnsymbol{footnote}}
M.Hotta
and M.Yoshimura\footnote[2]
{E-mail address: YOSHIM@tuhep.phys.tohoku.ac.jp\@.}\\
Department of Physics, Tohoku University\\
Sendai 980 Japan\\

\vspace{54pt}

{\bf Abstract}\\

\end{large}
\end{center}

\vsl
Quantum back reaction due to $N$ massless fields may be worked out to
a considerable detail in a variant of integrable dilaton gravity model
in two dimensions. It is shown that
there exists a critical mass of collapsing object of order
$\hbar N \times$ (cosmological constant)$^{1/2}$, above which the end point
of Hawking evaporation is two disconnected remnants of infinite
extent, each separated by a mouth from the outside region.
Deep inside the mouth there is a
universal flux of radiation in all directions, in a form
different from Hawking radiation. Below the critical
mass no remnant is left behind, implying complete Hawking evaporation
or even showing no sign of Hawking radiation.
Existence of infinitely many static states of quantum nature
is also demonstrated in this model.

\vspace{12pt}

\end{titlepage}

%
%

Although very little is known on the precise nature of quantum gravity,
it is quite plausible that a deep insight is gained by a thorough
understanding of the end point behavior of Hawking evaporation \cite{hw75}.
There has been a progress recently towards this goal in two dimensional
dilaton gravity theories in which the back reaction against Hawking radiation
may be taken into account in the form of a semiclassical approximation
\cite{cghs}, \cite{rstetc}.
The original model of Callan, Giddings, Harvey, and Strominger (CGHS),
however, suffers from the problem of curvature singularity, encountering the
breakdown of the semiclassical approach.

Most recently, we proposed a variant of dilaton gravity model \cite{hy93-2}
in which a classical solution describes the gravitational collapse of massive
body at rest and have shown that a wormhole is formed, radiating the black
body particles in much the same way as in the black hole radiation.
The crucial element here is  existence of the event horizon, not of the
singularity.
Consequently we studied the back reaction problem in this model, which
is guaranteed to be singularity-free. The model is however very difficult
to analyze in the original form in which the one loop matter effect is
included as the Polyakov action.

In this paper we modify the Polyakov action in such a way that the
semiclassical equations are solvable for massless fields.
This is the modification advocated by Russo, Susskind, and Thorlacius
(RST) \cite{rst2}. In a variant model of this type we shall be able to work
out and classify the end point behavior of Hawking
evaporation depending on the mass of the collapsing body.
There exists in this model a critical source mass of
$\sim 0.35\hbar N \times$ (cosmological constant)$^{1/2}$
that defines the boundary
for a complete evaporation of the hole. For larger masses the spacetime
at the end point consists of two disconnected worlds separated by two past
event horizons. Each disconnected world has a long throat of infinite
extent, separated by a mouth from the outside.
Inside the throat there is
a universal flux of radiation from all, in this case two, directions, which
however cannot be interpreted as the Hawking radiation.
This result is highly non-perturbative in both the mass and $\hbar$.
For very small masses even an approximate picture of Hawking
radiation loses its meaning due to the early onset of quantum back reaction.
In general, the Hawking radiation, even if it occurs, is a temporary and
a local phenomenon in this integrable model.

We furthermore establish existence of infinitely many static states above a
critical mass, $\frac{\hbar N}{12} \times$ (cosmological constant)$^{1/2}$.
How precisely these states are related to the problem of Hawking evaporation
is yet to be explored. But it is presumably an indication of infinitely
many quantum states bound by gravity that may store quantum information.

%
%

We consider the two dimensional dilaton gravity defined by the
classical action \cite{my92},
\begin{eqnarray}
S_{r} = \frac{1}{2\pi} \int\! d^{2} x \sqrt{-g}\;[\:e^{-2\varphi}
  \:( -R - 4\,\partial_{\mu}\varphi\,\partial^{\mu}\varphi + 4\lambda^{2})
+L^{(m)}\:]\,.
\end{eqnarray}
This model differs from the original CGHS model \cite{cghs},
in the two signs of the curvature and the dilaton kinetic terms.
The cosmological constant $\lambda^2 $ sets a length scale in the theory.
A novel feature of the dilaton gravity is that the coupling factor
$e^{2\varphi}$ acts as a varying gravitational constant,
as in the Brans-Dicke theory in four dimensions.
For the matter part $L^{(m)}$ we take a massive field as the source of
the gravity force.

It was shown in our previous paper \cite{hy93-2} that a unique wormhole
solution given by
\begin{eqnarray}
e^{-2\varphi}& =& e^{-2\lambda x^0 }
\sqrt{1-\frac{M}{2\lambda}\,e^{2\lambda (x^0 - |x^1 | )}\:},\\
e^{2\rho}&=&
\frac{1}{1-\frac{M}{2\lambda}
\, e^{2\lambda (x^0 - |x^1 | )}\,}\,,\label{whsol}
\end{eqnarray}
exists when a local massive source is put at $x^1 =0$, whose stress tensor
components are:
\begin{eqnarray}
T_{++} =T_{--} =T_{+-} = -\,\frac{M}{4\pi}\,e^{\rho}\,\delta (x^1 )\,.
\label{source}
\end{eqnarray}
In writing this solution we used the conformal coordinate,
\(\;
ds^{2}=\,-e^{2\rho}dx^{+}dx^{-}
\:\)
with
\(\:
x^{\pm}=x^{0}\pm x^{1}\,.
\:\)
The curvature is localized :
$
R = 2Me^{2\lambda x^0} \delta (x^1 )\,,
$
and the spacetime is everywhere flat except at the source.
The global event horizon is present at $x^- = x_H $ and $x^+ = x_H $ with
\(\:
x_H =-\,\frac{1}{2\lambda} \ln\frac{M}{2\lambda}\,.
\:\)

It is straightforward to analyze the classical spacetime of this wormhole
structure. Two patches of portions of flat spacetimes are glued at the
source. To a distant observer away from the source the massive body
appears to move with acceleration.
Furthermore it was shown \cite{hy93-2} by using the quantum stress tensor
obtained by one loop integration of $N$ massless fields, the Polyakov form,
that there is an outgoing flux of Hawking radiation from the source,
$\:\frac{N\lambda^2}{12\pi}\:$, in the asymptotic future infinity.

What happens to the spacetime structure if the back reaction due to Hawking
radiation is taken into account ?
This is the main theme of this paper.
We adopt the quantum
term in the way suggested by RST \cite{rst2} which has a merit of preserving
a classical symmetry even at the quantum level: the crucial key to the
integrability.

The RST modification is defined by the effective quantum action $S_q $
to be added to the classical $S_r $ of the form,
\begin{eqnarray}
S_{q} =\:-\,\frac{\kappa}{8\pi} \int\! d^{2} x\:\sqrt{-g}\:
[\: R\frac{1}{\Box}R+2\,\varphi R\:]\,,
\end{eqnarray}
with $\:\kappa = \frac{\hbar N}{12}\:$ for the large $N$ limit.
The basic equations in this reversed RST model may then be written
in the conformal gauge as follows:
\begin{eqnarray}
\partial_+ \partial_- (e^{-2\varphi} +\frac{\kappa}{2} \varphi -\kappa \rho)
 -\lambda^2 e^{2\rho -2\varphi} &=& \pi T_{+-}\,, \label{eqn1}\\
(1+\frac{\kappa}{4} e^{2\varphi} )\,\partial_+ \partial_- (\rho-\varphi)
&=& -\,\frac{\pi}{2} e^{2\varphi}\,T_{+-}, \label{eqn2}\\
 \partial_{\pm}^2 \,(e^{-2\varphi} +\frac{\kappa}{2}\varphi -\kappa\rho)
+\partial_{\pm}(\rho-\varphi)\, \partial_{\pm} (-2 e^{-2\varphi} +\kappa\rho)
&=& \kappa t_{\pm} -\pi T_{\pm\pm}\,,
\label{eqn3}
\end{eqnarray}
where $t_{\pm}$ is the boundary term specified by no incoming flux
condition at past infinity.
The effective gravitational constant $G_{eff}$ defined by
\(\:
R = 2\pi\,G_{eff}T_{\mu}^{\mu}\, +
\)
(dilaton and cosmological constant terms) is given by
\(\:
G_{eff} = e^{2\varphi}/(1+\frac{\kappa}{4}\,e^{2\varphi})^2  \,.
\:\)
Thus both in the weak and the strong coupling limit, $\varphi = \pm\infty$,
$G_{eff}$ is small.
Off the local source this semiclassical model is solvable even with
presence of the massless fields \cite{hsty}. In our problem at hand we
must furthermore
impose the junction condition derived by integrating the above equations for
an infinitesimal interval around the source,
which relates discontinuity
of derivative of fields at the source to the mass $M$.

%
%

Let us briefly describe how to derive the quantum solution.
Start with the general solution \cite{hsty} off the source.
This solution involves two sets of free fields and the constraint equation
is first solved in favor of one set of the free fields.
We then impose the boundary condition:  the solution asymptotically
approaches the classical wormhole solution at infinite past. By a choice
of an appropriate coordinate gauge the boundary term is fixed by
$t_{\pm} = 0$. The junction condition is then reduced to a pair of
integro-differential equations to be solved for a single variable function,
involving the dilaton field at the source  as a parameter,
which will be written
later in an equivalent form.
The solution thus derived is symmetrical with respect to the source, hence
one may write it only in the right region, $x^1 > 0 $. The solution in the
other region, $x^1 < 0$, is obtained by the replacement, $x^+
\leftrightarrow x^- $.
In terms of the yet to be determined function
$F_- $, the quantum solution with the specified boundary condition is
given by
\begin{eqnarray}
e^{-2\varphi}-\frac{\kappa}{2}\,\varphi &=&
e^{-\lambda x^{+}}\,F_{-}(x^{-})-\frac{\kappa}{4}\,\lambda\,(x^+ -x^- )
+\frac{\kappa}{2}\,\ln\,|F_{-}'(x^- )/\lambda|+\frac{\kappa}{4}
\nonumber \\
&&
+\frac{\kappa}{4}\,F_{-}\int_{-\infty}^{x^- }\,dx\,
\frac{(F_{-}'')^2 }{(F_{-}')^3 }
-\frac{\kappa}{4}\,\int_{-\infty}^{x^- }\,dx\,[\:F_{-}
\frac{(F_{-}'')^2 }{(F_{-}')^3 }+\lambda\:]\,,\\
e^{2(\rho-\varphi)} &=&  -\,\frac{1}{\lambda}\,e^{-\lambda x^+ }
F_{-}'(x^- )\,.
\end{eqnarray}

A general analysis shows that consistency with the junction condition
gives no turning point for $F_{-}\::$ $F_{-}'\neq 0$ at any finite $x^{-}$.
Furthermore the asymptotic behavior at past is the classical one,
\(\:
F_- ^{cl} = \sqrt{e^{-2\lambda\,x }-\frac{M}{2\lambda}\,}
\:,\)
and it is a decreasing function of the argument $x$.
Hence, $F_{-}$ monotonically decreases for the quantum case as well:
$F_- ' <0 $.

It turns out that it is much more convenient to anlayze the ordinary
differential equation derived from the above integro-differential equation
expressing the junction condition.
In terms of the following single variable functions,
\begin{eqnarray}
y=(-\,F_{-}'/\lambda)^{-\frac{1}{2}\,}\,e^{\frac{\lambda}{2}\, x}\,,
\hspace{0.5cm} g=e^{\bar{\varphi}}\,, \hspace{0.5cm}
h=\frac{1}{\lambda}\,g'\,,
\end{eqnarray}
with $\bar{\varphi} = \varphi(x^1 = 0)$,
the equations to be analyzed are
\begin{eqnarray}
\frac{1}{\lambda}\, g' &=& h \,, \hspace{0.5cm}
\frac{1}{\lambda}\,y' = y-\frac{1}{4}\,\frac{ag^3 }{1+\frac{\kappa}{4} \,
g^2 }\,,\\
\frac{1}{\lambda}\,h' &=& -\,h+\frac{1}{4}\,\frac{ag^3 }
{1+\frac{\kappa}{4} \,g^2 }\,
\frac{h}{y}\,[\,1-\frac{\kappa}{2}\,\frac{g^2 (3+\frac{\kappa}{4}\,
g^2 )}{(1+\frac{\kappa}{4} \,g^2 )^2 }\,]
+\frac{3+\frac{\kappa}{4}\,g^2 }{g\,(1+\frac{\kappa}{4} \,g^2 )}\,h^2
\nonumber \\ &&
+\frac{1}{4y}\,\frac{g^3 }{1+\frac{\kappa}{4} \,g^2 }
[\:\frac{1}{4y}\,\frac{a^2 g^4 }{1+\frac{\kappa}{4} \,g^2 }
-\frac{4}{y}+\kappa y-\frac{\kappa}{8y}\,
\frac{a^2 g^6 }{(1+\frac{\kappa}{4} \,g^2 )^2 \,}\:]\,,
\end{eqnarray}
with $a= M/\lambda \,.$

%
%

The set of differential equations we thus derived
is readily amenable to numerical
analysis, but it is crucial to have an analytic control of late time
behaviors for a deeper understanding.
In order to classify the end point behavior of solutions, it is important
first to enumerate fixed points of this set of equations.
There are two fixed points:
(1) $(g\,, h\,, y)=(\infty\,,\infty\,,\infty)\:$,
(2) $(g\,, h\,, y)= (2\,(a-\kappa)^{-1/2}\,,
0\,, 2\,(a-\kappa)^{-1/2})\:$ for $M > \kappa\lambda$.
It turns out that the second fixed point has one direction attractive,
but the other two directions repulsive, hence it is irrelevant to
discussion of the end point behavior.
But it will later be shown that the second fixed point has another profound
implication; it is related to existence of infinitely many static states.
On the other hand, the first fixed point is an attracter and
it describes the late time behavior of evaporating hole:
\(\,
g= O[\,e^{\beta\lambda x }\,] \,, \hspace{0.5cm}
y = O[\,e^{\beta\lambda x }\,] \,,\)
with $0 < \beta < 1$ at $x = \infty$.
There is a relation between these constants:
\(\:
\beta = 1-\frac{M}{\kappa\lambda}\,\lim_{x \rightarrow \infty}\,(g/y)
\:.\)
The curvature calculated in this case yields a limiting behavior of
$R = O[\,e^{-2\beta\lambda x^0 }\,]$ at late times.
The coupling behaves like $\varphi \sim \beta\lambda x^0 $.
Thus this case describes the final flat spacetime without event horizon.

Yet another important end point behavior is an approach to the allowed
boundary, $y = 0 $. As both verified by numerical analysis and the following
analytic method, it is possible to have solutions approaching
this boundary at a finite $x = x_c $ according to
\begin{eqnarray}
y &\sim& y_0 (x_c -x )-y_1 (x_c -x )^{\alpha+2}\,, \\
g &\sim& g_c -\frac{h_0 }{\alpha+1}\,(x_c -x )^{\alpha+1}\,,
\end{eqnarray}
with
$-1<\alpha<0$.
Consistency leads to a critical mass given by
$M_{cr} \simeq 4.2\,\kappa\lambda\,,$
above which this behavior is realized when the parameters $g_c $ and $y_0 $
are given in terms of $a = M/\lambda $ by
\begin{eqnarray}
 \frac{\kappa}{4}\,a^2 g_c ^6 -(a^2 -\kappa^2 )g_c ^4
+8\kappa g_c ^2 +16 =0\,,
\end{eqnarray}
and
\(\:
 y_0 = \frac{\lambda}{4}\,ag_c ^3 /(1+\frac{\kappa}{4} \,g_c ^2 )\,.
\:\)
The exponent $\alpha$ obeys:
$ \alpha^2 +B\alpha -C=0\,,$ with $B(M/\lambda)\,,$ and $ C(M/\lambda)$
functions of the mass. A real solution of the exponent $\alpha$
with $-1<\alpha<0$ exists only for
$M > M_{cr}$. There is also a relation between the coefficients of the next
leading terms, $y_1 $ and $h_0 $.

\begin{center}
\leavevmode
\epsfysize=6cm \epsfbox{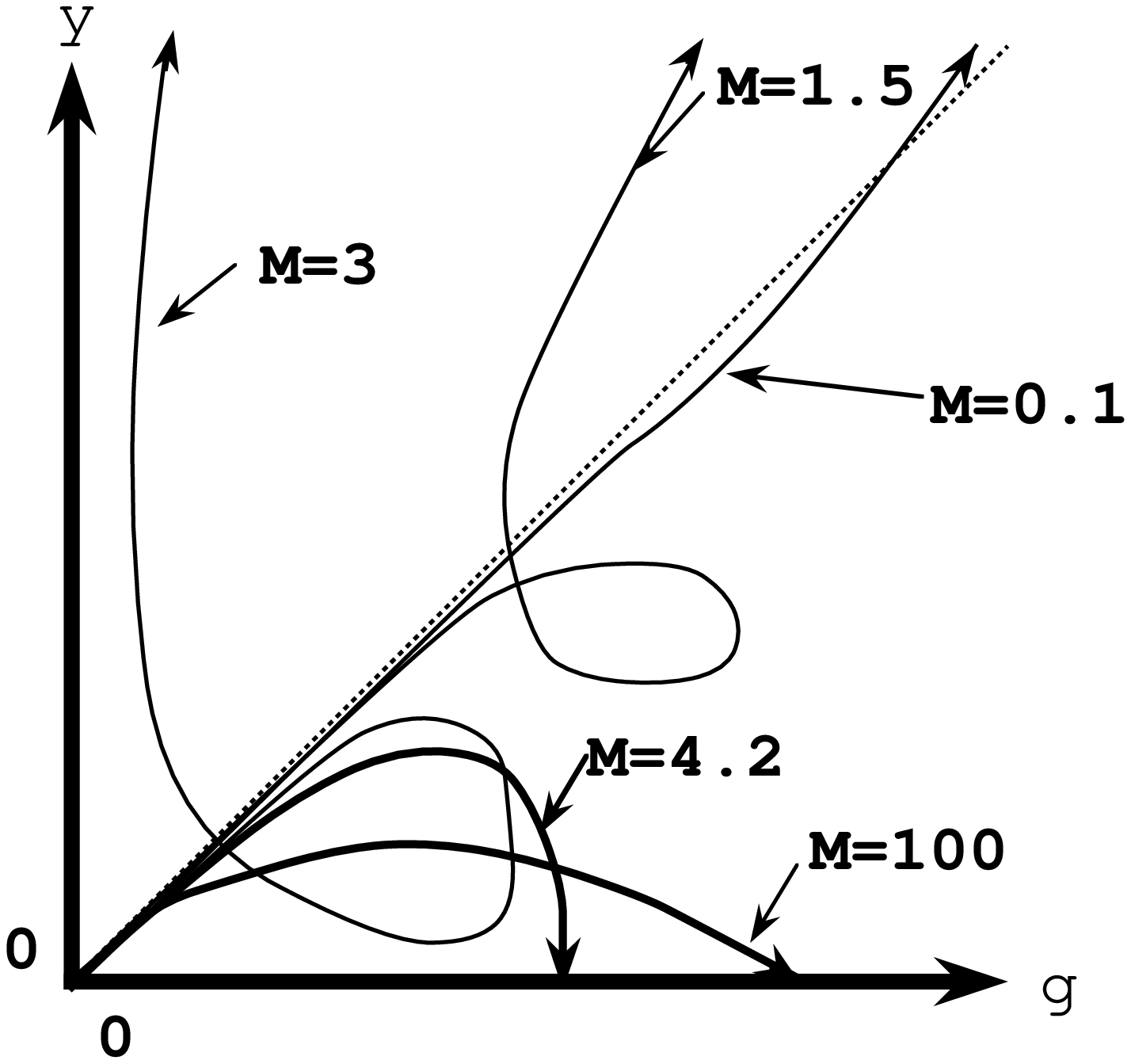}
\end{center}
\begin{center}
{\parbox{13cm}{\small  Fig 1.
Schematic trajectories of the dilaton field $g = e^{\varphi} $
and $y = (-\,F_- '/\lambda)^{-1/2}\,e^{\lambda x/2} $ at the source
for various masses, $M $ measured in unit of $\kappa\lambda$.
Trajectories start at the origin with the classical behavior,
\(\:
y = -\,\frac{a}{4}\,g^3 +\sqrt{g^2 + \frac{a^2 }{16}\,g^6 } \,,
\; a = M/\lambda \,,
\:\)
and evolve as the arrows indicate.}}
\end{center}

In Fig.1 is plotted the schematic outline of trajectories projected onto
the $(g\,, y)$ plane for various source masses. It is seen that
above the critical mass of $ \sim 4.2\kappa\lambda$ the
coupling at the source $g$ approaches a finite value,
while below the critical mass it goes to infinity in the end,
exhibiting complicated transient behaviors at intermediate masses.

Before discussing meaning of these results, let us enumerate the limiting
behaviors of the parameters $g_c \,, y_0 \,, $ and $\alpha$:
the large mass limit $M \rightarrow \infty$ is given by
\(\:
g_c ^2 \sim
\frac{4}{\kappa}\,[1-\frac{\kappa^2 \lambda^2 }
{M^2 }]\,, \hspace{0.5cm} y_0 \sim \frac{1}{\kappa\sqrt{\kappa}\,}\,M\,,
\hspace{0.5cm} \alpha \sim -2\,\frac{\kappa\lambda}{M}\,,
\:\)
while the critical mass limit $M \rightarrow M_{cr}$ is by
\(\:
 g_c ^2 \sim \frac{2.2}{\kappa}\,[1 + 1.1\delta/\sqrt{\kappa}\,] \,,
 \hspace{0.5cm}  \alpha+1 \sim \delta/0.79 \,,
\:\)
with
\(\:
\delta = \sqrt{\frac{M-M_{cr}}{M_{cr}}}\,.
\:\)

%
%

Let us now discuss the $M>M_{cr}$ case. It is straightforward to write down
the leading end point behavior near $x^- \sim x_c $ in the region right to
the source,  $x^+ > x_c $;
\begin{eqnarray}
e^{-2\varphi}-\frac{\kappa}{2}\,\varphi &\sim&
\frac{\lambda}{y_0 ^2 }\,\frac{1}{x_c -x^-  }\,
[1-e^{-\lambda(x^+ -x_c )\,}]
-\frac{\kappa}{4}\,\lambda x^+ \,,\\
 e^{2(\rho-\varphi)} &\sim& \frac{1}{y_0 ^2 }\,e^{-\lambda(x^+ -x_c )}
\frac{1}{(x_c -x^- )^2 }\,.
\end{eqnarray}
For a finite $x^+ $ the dilaton field $\varphi$ approaches a negative
infinity as $x^- \rightarrow x_c ^- $: a weak coupling behavior,
and the metric
$e^{2\rho}$ has a factorized form of two separate functions of $x^{+}$
and $x^{-}$.
It is thus clear that the spacetime at $x^- \sim x_c $ with a finite
$x^+ $ is almost flat, because $R \propto \partial_+ \partial_- \rho =0 $.
Moreover, since $e^{-2\varphi} \propto 1/(x_c -x^- ) $,
the metric $e^{2\rho} \propto 1/(x_c -x^- ) $.
This behavior is characteristic of the global event horizon. Indeed,
the flat coordinate, $ds^2 \sim -\,d\xi^+ d\xi^- $,
in this portion of the spacetime is given by
\begin{eqnarray}
\xi^+ = \frac{1}{\lambda}\, \ln (1-e^{-\lambda(x^+ -x_c )}) \,, \hspace{0.5cm}
\xi^- = -\,\frac{1}{\lambda}\, \ln \lambda(x_c -x^- ) \,, \nonumber
\end{eqnarray}
with an infinite range of
$\: -\infty<\xi^+ < 0 \:, -\infty < \xi^- < \infty \,.$
It turns out that the two lines of $x^{+} = x_c\,, x^1 > 0 \,$ and
$x^{-} = x_c\,, x^1 < 0 \,$
are the past event horizon as in the classical case.

The curvature in the vicinity of the local source may be evaluated
by directly computing $R = 8e^{-2\rho}\,\partial_+ \partial_- \rho$.
Its end point behavior is then given by
\(\:
R \sim -\,4\lambda^2\,(1+\frac{16}{a^2 g_c ^4 })
/(1+\frac{\kappa}{4}\,g_c ^2 )\,
\,,
\:\)
having a large mass limit of
$\sim -\,2\lambda^2 $ for $M \rightarrow \infty$.
At the source an infinite proper time is needed to approach the end point,
\begin{eqnarray}
\tau = \int^{x_c } \,dx^0 \,e^{\bar{\rho}} \sim
\frac{g_c }{y_0 }\,\int^{x_c }\, \frac{dx^0 }{x_c -x^0 } = \infty \,,
\end{eqnarray}
in contrast to a finite value in the classical case.

How about the other side of the end point, $x^+ = \infty$ ?
This portion of the spacetime is in the strong coupling region;
\(\:
\varphi \sim \frac{1}{2}\,(\lambda x^+ -D_- )\,,
\:\)
with
\(\:
D_- \propto 1/(x_c -x^- ) \,,
\:\)
near $x^- = x_c $.
The metric behaves as
\(\:
e^{2\rho} \propto e^{-D_- }/(x_c -x^- )^2 \,.
\:\)
It can be shown that the curvature
$R = O[e^{-\lambda x^+ }]\,$ in this region.
This flat region may be called the outside, and the previous $x^- = x_c $
region the inside or the throat. The long throat is separated by the
outside at the corner $(\,x^- = x_c \:, x^+ = \infty \,)$, which may be
called the mouth.

%
%

Information on the Hawking radiation is contained in the quantum
stress tensor components derived by one loop matter integration, which
is given in the reversed RST model as
\begin{eqnarray}
&&
\langle T_{\pm\pm} \rangle = -\,\frac{\kappa}{\pi}\,\left[\,
\partial_{\pm}^{2}\rho - (\partial_{\pm}\rho)^{2} +
\partial_{\pm}\rho\,\partial_{\pm}\varphi
-\frac{1}{2}\,\partial_{\pm}^{2}\varphi \right]\,,\label{qfluxrst2} \\
&&
\langle T_{+-} \rangle = \,\frac{\kappa}{2\pi}\,\partial_{+}\partial_{-}
(2\rho-\varphi)\,.\label{qfluxrst1}
\end{eqnarray}
Note that this formula is valid only in the original $x^{\pm}$ coordinate,
since the boundary term $t_{\pm}(x^{\pm})$ vanishes only in this coordinate
and it is not a tensor. Interpretation of the stress component is however
clearest in the locally flat coordinate system, hence the coordinate
transformation must be performed.

With this remark in mind one may show
that the asymptotic flux at future null infinity $x^+ = \infty$ always
vanishes. First, above the critical mass, $M > M_{cr} $, one expresses the
stress tensor in terms of $g, y, h$ by using the relevant form,
\(\:
\varphi \sim \frac{1}{2}\,(\lambda x^+ -D_- )
\:.\)
By using the exact differential equations for these quantities, one can
rigorously prove that
\(\:
\langle T_{\pm\,\pm} \rangle = 0 = \langle T_{+-} \rangle
\:.\)
The vanishing stress tensor in the subcritical case is much easier to
demonstrate.

Does this result mean that the Hawking radiation never exists ?
We do not think that this is necessarily so.
It only seems to imply that
Hawking radiation is a temporary and a local phenomenon,
not a universal one
independent of any observer far away from the collapsing body.
But one may certainly
have a situation in which the Hawking process lasts for a large time interval.
To illustrate this point, consider an observer at a finite $x^1 $.
One can define a region in which the semiclassical picture is valid by
comparing the two terms in the determinental equation;
\begin{eqnarray}
e^{-2\varphi}-\frac{\kappa}{2}\,\varphi =
e^{-\lambda x^+ }\,F_- -\frac{\kappa}{4}\,(\lambda x^+ -D_- )\,.
\end{eqnarray}
The semiclassical region is given by
\(\:
e^{-\lambda x^+ }\,|F_-| \geq |\,\frac{\kappa}{4}\,(\lambda x^+ -D_- )\,|
\:.\)
If the Hawking picture is valid in this region, this inequality must be
obeyed even when one inserts the classical solution near the event horizon;
\begin{eqnarray}
e^{-2\lambda x^0 }\,\sqrt{2\lambda(x_H -x^- )\,} \gg
\frac{\kappa}{2}\,|\,\lambda x^0 + \frac{3}{4}\,\ln 2\lambda(x_H -x^- )\,|\,.
\nonumber
\end{eqnarray}
This yields a limited region of
$\, |\lambda x^1 | < O[\, \ln \frac{M}{\kappa\lambda} \,] \:$
where one can see the Hawking radiation clearly.
The duration for which it lasts is of $O[1/\lambda]$ around
$x^0 = x_H +x^1 $.

The behavior of the stress tensor near the end point, $x^- = x_c $,
is also interesting. One finds
for the stress components in the flat coordinate $\xi^{\pm}$,
\begin{eqnarray}
&& \langle T_{\xi^- \xi^- } \rangle  = (\frac{dx^- }{d\xi^- })^2\,
\langle T_{--} \rangle \sim -\,\frac{\kappa\lambda^2 }{4\pi} \,,\\
&& \langle T_{\xi^+ \xi^+ } \rangle  \sim -\,\frac{\kappa\lambda^2 }{4\pi}\,
\frac{1-2e^{\lambda\xi^+ }}{(1-e^{\lambda\xi^+ })^2 \,} \,.
\end{eqnarray}
Away from the mouth region, $|\lambda \xi^+ | \gg 1 $, and
there exists radiation flux of both right and left movers of equal magnitude,
$\frac{\kappa\lambda^2 }{4\pi} = \frac{N\lambda^2 }{48\pi}$.
This value coincides with the universal flux observed in the final state of
the cosmological model \cite{hsty},
but differs from the semiclassical flux radiating from the classical event
horizon by a factor $\frac{1}{4}$. As one approaches the mouth region,
$\xi^+ \rightarrow 0$, and the left moving component of the stress tensor,
$\langle T_{\xi^+ \xi^+ } \rangle$, behaves strangely: it changes the sign
and tends to diverge. But this is precisely the region in which the above
approximation breaks down.

%
%

Finally, we demonstrate existence of infinitely many static states above
a critical mass, $M>\kappa\lambda$. Reconsider the second case of the
fixed point. Perturbation around this fixed point indicates
that the unique time dependent approach to the
fixed point is obtained with
$y = 2\,(a-\kappa)^{-1/2} $
fixed and
$ 2ag'+\frac{1}{\lambda}\,g'' = 0 \,, $ giving
\(\:
g = 2\,(a-\kappa)^{-1/2}+ \epsilon\, e^{-2Mx} \,,
\:\)
with $\epsilon$ an infinitesimally small parameter.

Spacetime structure of this solution is derived from
\(\:
-\,F_- '/\lambda = e^{\lambda x}/y^2  =
\frac{1}{4}\,(a-\kappa)\,e^{\lambda x} \,.
\:\)
Asymptotically as $x^0 \rightarrow \infty$,
\begin{eqnarray}
 e^{-2\varphi}-\frac{\kappa}{2}\,\varphi &=&
-\,\frac{1}{4}\,(a-\kappa)\,e^{-2\lambda |x^1 |} - \frac{\kappa}{2}\,\lambda
|x^1 | + \kappa C
\,, \\
 e^{2\rho} &=& e^{2\varphi}\,\frac{1}{4}\,(a-\kappa)\,e^{-2\lambda |x^1 |}\,.
\end{eqnarray}
The solution thus represents a localized static state.
The calculated curvature is everywhere regular and negative, implying
that this state acts as a source of attraction. The asymptotic behavior
at $|x^1 |= \infty$ is like
\(\:
R = -\,O[\,\lambda^2 /(\kappa(a-\kappa))\,e^{-2\lambda|x^1 |}\,] \,,
\:\)
hence it has no classical analogue at $\kappa=0$.
Moreover a nontrivial node with $\partial_1\,\varphi = 0$ exists at
\(\:
|x^1 | = \frac{1}{2\lambda}\,\ln (\frac{M}{\kappa\lambda}-1) \,,
\:\)
for $M > 2\kappa\lambda$.

The solution thus derived is valid in the asymptotic future region, because
we can only analyze small deviation away from the fixed point. But
existence of a global time dependent solution approaching this asymptotic
behavior should be evident from the uniqueness of the solution to the
differential equations.
Furthermore it should not be confused that this solution has
direct relevance to the gravitational collapse of mass $M$ considered up to
now, since the past boundary condition, though not known at present,
is expected to be different from the classical wormhole condition.
But some kind of perturbation to the standard collapse model should be able
to push the configuration towards the fixed point.
We may then expect that
this static state can appear in the collision process, for instance,
a collision between the hole and a massless shock wave.
This does not however mean that it is easy to excite the static state
in the collision process, since for a specified mass $M$
there is only one unique trajectory
ending at the fixed point, which suggests that the perturbation
must be fine tuned to excite this state with a large cross section.
Existence of static states for
any mass above the critical one is presumably an indication of infinitely many
quantum objects yet to identified.
The quantum information apparently lost by emission of the black body radiat
ion
may be stored in these objects.
We hope to cover these problems in future.

%
%

Our model is only a toy model of Hawking evaporation and should not be
taken too seriously. But it incorporates many novel ideas that have been
proposed so far to resolve the information loss paradox;
creation of a baby universe \cite{dyson}, \cite{whinf}, quantum
remnant in the form of horned particles \cite{banks92-1},
existence of infinitely degenerate quantum states \cite{acn87},
and early onset of the back reaction \cite{bhevaporev}.
It is surprising that fragments of these ideas are already available
on the market. Perhaps our sole contribution here is to have a coherent and
an integrated picture for these ideas in the form of the integrable model.
This, we believe, is more than enough to justify our insistence on a
particular model.

In summary, we constructed a model of the gravitational collapse of a massive
body and pursued its fate caused by quantum back reaction due to $N$
massless fields.
There is a critical mass of the body of order
$4.2\times \frac{\hbar N}{12}\,\times$ (cosmological constant)$^{1/2}$.
Above this mass the end point of the collapse is a pair of remnants
surrounded by the quantum
mouth. Inside the mouth there is a universal flux of radiation
everywhere in a form different from Hawking radiation.
Below the critical mass the Hawking evaporation cannot last for long,
producing a flat spacetime without horizon in the end.
The Hawking process is a transient and a local
phenomenon, not universal to any distant observer.
Furthermore existence of infinitely many static states is demonstrated for
masses above $M > \frac{\hbar N}{12}\,\times$ (cosmological constant)$^{1/2}$.


\newpage

\begin{thebibliography}{99}

%
\def\AP{{\sl Ann.\ Phys.\ {\rm(}N.Y.{\rm)} }}
\def\CMP{{\sl Commun.\ Math.\ Phys.\ }}
\def\FP{{\sl Fortsch.\ Phys.\ }}
\def\GRG{{\sl Gen.\ Rel.\ Grav.\ }}
\def\JMP{{\sl J.\ Math.\ Phys.\ }}
\def\JPSJ{{\sl J.\ Phys.\ Soc.\ Jpn.\ }}
\def\LNC{{\sl Lett.\ Nuovo Cim.\ }}
\def\LNCI{{\sl Lett.\ Nuovo Cim.\ }(Ser.~I), }
\def\MPL{{\sl Mod.\ Phys.\ Lett.\ }}
\def\MPLA{{\sl Mod.\ Phys.\ Lett.\ A }}
\def\NC{{\sl Nuovo Cimento }}
\def\NCA{{\sl Nuovo Cimento A }}
\def\NP{{\sl Nucl.\ Phys.\ }}
\def\NPB{{\sl Nucl.\ Phys.\ B }}
\def\PL{{\sl Phys.\ Lett.\ }}
\def\PLA{{\sl Phys.\ Lett.\ A }}
\def\PLB{{\sl Phys.\ Lett.\ B }}
\def\PRep{{\sl Phys.\ Rep.\ }} \let\PREP=\PRep
\def\PR{{\sl Phys.\ Rev. }}
\def\PRA{{\sl Phys.\ Rev.\ A }}
\def\PRD{{\sl Phys.\ Rev.\ D }}
\def\PRL{{\sl Phys.\ Rev.\ Lett.\ }}
\def\PS{{\sl Physica Scripta }}
\def\PTP{{\sl Prog.\ Theor.\ Phys.\ }}
\def\PTPS{{\sl Prog.\ Theor.\ Phys.\ Suppl.\ }}
\def\ZPC{{\sl Z.\ Phys.\ C\ }}
%

\bibitem{hw75} S.W.Hawking, \CMP{\bf 43}, 199(1975).
\bibitem{cghs} C.G.Callan, S.B.Giddings, J.A.Harvey and A.Strominger,
\PR,{\bf D45},\newline 1005(1992).
\bibitem{rstetc} J.G.Russo, L.Susskind and L.Thorlacius,
\PL{\bf 292B},13(1992);

T.Banks, A.Dabholkar, M.R.Douglas and M.O'Loughlin,
\PR{\bf D45}, 3607(1992);

S.W.Hawking, \PRL{\bf 69}, 406(1992);

B.Birnir, S.B.Giddings, J.A.Harvey, and A.Strominger,
\PR{\bf D46}, 638(1992);

L.Susskind and L.Thorlacius, \NP{\bf B382}, 123(1992).

\bibitem{hy93-2} M.Hotta and M.Yoshimura, {\it Wormhole and Hawking Radiatio
n},
Tohoku University preprint, TU/93/440.

\bibitem{rst2} J.G.Russo, L.Susskind and L.Thorlacius,
\PR{\bf D46}, 3444(1992).
\bibitem{my92} M.Yoshimura, \PR{\bf D47}, 5389(1993).
\bibitem{hsty} M.Hotta, Y.Suzuki, Y.Tamiya, and M.Yoshimura,
\PR{\bf D48}, 707(1993);

M.Hotta, Y.Suzuki, Y.Tamiya, and M.Yoshimura, {\it Universality of Final State
in Two Dimensional Dilaton Cosmology}, \PTP,{\bf 90}, No.3 in press
(September, 1993).

\bibitem{dyson} F.Dyson, unpublished (1976).
\bibitem{whinf}
S.W.Hawking, \PR{\bf D37}, 904(1988);
and references therein.

\bibitem{banks92-1}
T.Banks, A.Dabholkar, M.R.Douglas and M.O'Loughlin,
\PR{\bf D45}, 3607(1992);

T.Banks and M.O'Loughlin, \PR{D47}, 540(1993).

\bibitem{acn87} Y.Aharonov, A.Casher, and S.Nussinov, \PL{\bf 191B},
51(1987).

\bibitem{bhevaporev} For a recent review,
J.Preskill, {\it Do Black Holes Destroy Information ?},
Caltech preprint, CALT-68-1819, (1992).


\end{thebibliography}
\end{document}